\documentclass[9pt,twocolumn,twoside]{pnas-new}

\templatetype{pnasresearcharticle} 

\newcommand{\rr}{  {\mathbf r} }
\newcommand{\ee}{ {\mathbf e} }

\newcommand{\p}{\partial}
\newcommand{\BEQ}{\begin{equation}}
\newcommand{\EEQ}{\end{equation}}
\newcommand{\BEA}{\begin{eqnarray}}
\newcommand{\EEA}{\end{eqnarray}}

\newcommand{\mtt}[1]{\textcolor{black}{#1}}

\title{Probing the non-Debye low frequency excitations in glasses through random pinning}

\author[a,b]{Luca Angelani}
\author[c]{Matteo Paoluzzi} 
\author[b,d,e]{Giorgio Parisi}
\author[b,f]{Giancarlo Ruocco}

\affil[a]{ISC-CNR,  Institute  for  Complex  Systems,  Piazzale  A.  Moro  2,  I-00185  Roma,  Italy}
\affil[b]{Dipartimento di Fisica, Sapienza Universit\`a di Roma, Piazzale A. Moro 2, I-00185, Rome, Italy}
\affil[c]{Department of Physics and Syracuse Soft \& Living Matter Program, Syracuse University, Syracuse NY 13244, USA}
\affil[d]{Nanotec-CNR, UOS Rome, Sapienza Universit\`a di Roma, Piazzale A. Moro 2, I-00185, Rome, Italy}
\affil[e]{INFN-Sezione di Roma 1, Piazzale A. Moro 2, 00185, Rome, Italy}
\affil[f]{Center for Life Nano Science, Istituto Italiano di Tecnologia, Viale Regina Elena 291, I-00161, Rome, Italy}

\leadauthor{Lead author last name} 

\significancestatement{
Amorphous solids are continuum media. However, their
mechanical and thermodynamical properties, even though 
universal, dramatically deviate from those in crystalline solids.
Their anomalous behavior reflects peculiar and universal deviations from Debye's law in the low-frequency sector of the density of states $D(\omega)$. 
Theoretical models predict a population
of non-Goldstone bosons following the universal power
law $D(\omega)\sim \omega^4$ that are subdominant and
then hard to detect. In this work, we introduce
a general protocol 
that can be employed in both,
numerical simulations and experiments, to probe
the non-Debye portion of the spectrum.}

\authorcontributions{Author contributions: G.P. conceived research; L.A., M.P., G.P., and G.R. designed research; M.P. and G.R. performed research; M.P. wrote the paper.}
\authordeclaration{Please declare any conflict of interest here.}
\equalauthors{\textsuperscript{1}A.O.(Author One) and A.T. (Author Two) contributed equally to this work (remove if not applicable).}
\correspondingauthor{\textsuperscript{2}To whom correspondence should be addressed. E-mail: author.two\@email.com}

\keywords{glasses $|$ normal modes $|$ non-Debye law $|$ soft modes} 

\begin{abstract}
We investigate the properties of the low-frequency spectrum in the density of states $D(\omega)$ 
of a three dimensional {\color{black} model} glass former. To magnify the Non-Debye sector of the spectrum, we introduce
a random pinning field that freezes a finite
particle fraction {\color{black} in order to 
break the translational invariance and shifts 
all the vibrational frequencies of the extended modes towards higher frequencies}  . We show that Non-Debye soft localized modes progressively emerge as
the fraction $p$ of pinned particles increases. 
Moreover, the low-frequency tail of $D(\omega)$ 
goes to zero as a power law $\omega^{\delta(p)}$, with $2 \!\leq \! \delta(p) \!\leq\!4$ and $\delta\!=\!4$ above a threshold fraction $p_{th}$. 
\end{abstract}

\dates{\today}
\doi{\url{www.pnas.org/cgi/doi/10.1073/pnas.XXXXXXXXXX}}

\begin{document}

\verticaladjustment{-2pt}

\maketitle
\thispagestyle{firststyle}
\ifthenelse{\boolean{shortarticle}}{\ifthenelse{\boolean{singlecolumn}}{\abscontentformatted}{\abscontent}}{}

\dropcap{U}nderstanding the peculiarities and the universal features of the 
low-frequency spectrum in glasses plays a crucial role to gain
insight into their thermal and mechanical properties.

In the case of crystalline solids, 
mechanical and thermal properties  
follow universal laws that can be 
obtained through Debye's theory. Debye's law assumes that 
the only energy excitations around the ground state in crystals 
are phonons, i. e., Goldstone modes. 
The corresponding density of sates $D(\omega)$ 
below Debye's frequency follows $D(\omega)\sim \omega^{d-1}$,
in $d$ spatial dimensions \cite{kittel2005introduction}.

More complex is the situation for amorphous systems where the 
low-frequency spectrum shows an abundance of soft non-Goldstone modes. Quasilocalized 
soft modes are involved, for example, in the relaxation processes of
supercooled liquid \cite{widmer2008irreversible}, and in the plastic flow of disordered
solids \cite{manning2011vibrational}. They have a natural interpretation within
the jamming transition. 
At the jamming point, a glass is isostatic, i. e., the number of degrees of
freedom equals the number of constraints \cite{Hern,Wyart12}.
This means that soft modes are induced as soon as a particle contact is removed.  
More in general, thanks to a class of exactly 
solvable mean-field models for which it has been proven that the energy landscape is organized into a complex hierarchy  of marginally stable states \cite{Franz15,kurchan2013exact,charbonneau2017glass}, 
marginal stability has been suggested as a general ingredient for the rising of no-energy
cost excitations connecting degenerate minima that are separated by small energy barriers \citep{mezard84,charbonneau2014fractal}.

Theoretical models of random media predict a universal law for the density of 
the states of the non-Goldstone {\color{black} (i.e. non-phononic)} component of the 
spectrum with a scaling $D(\omega)\sim \omega^4$ in any dimensions 
\cite{Gurarie03,Gurevich}. {\color{black} Consistently, it has been recently shown that in 
many real and simulated glasses the non-Debye contribution to the density of states
is proportional to the phonon damping, thus showing the well know Reyleigh fourth power 
frequency dependence \cite{Schirmacher2007}}. However, since on large enough scale 
glasses are continuum media, the phononic contribution in $D(\omega)$ dramatically 
overcomes any subdominant non-Goldstone
tail at low frequencies \cite{tanaka}. \mtt{Numerical simulations of repulsive binary
mixture suggests that the Goldstone modes hybridizes with non-Goldstone excitations and
destroys the $\omega^4$ tail \cite{lerner2016nonlinear}.
}

Recently, it has been possible to observe a 
non-Goldstone  low-frequency 
sector of the spectrum in the density of states of 
structural glasses and disordered systems that follows a power law $D(\omega)\sim \omega^4$\cite{Baity-Jesi15,Lerner2016,Lerner_rapid,lerner2016statistics,Mizuno14112017}. 
In order to take access to non-Debye modes in numerical models, 
one has to perform simulations of extremely large system sizes \cite{Mizuno14112017}. Alternatively, one has
to find a protocol to {\color{black} cancel}  the Goldstone bosons from the low-frequency region \cite{Lerner2016}
{\color{black} by choosing suitably small systems}. 

In the present paper,
{\color{black} in analogy with the procedure adopted} 
in Ref. \cite{Baity-Jesi15} where a random field
has been introduced in the Heisenberg spin-glass to destroy the spin-wave 
contribution at low frequencies, 
we employ a random pinning field that freezes a finite fraction of particles. 
The presence of this random external field destroys any spatial symmetry, 
removing the corresponding Goldstone excitations. 

Random pinning has been largely employed to gain more information
about Random-First-Order Theory in glass-forming liquids \cite{kob2012non_pin,cammarota2012ideal_pin,ozawa2015equilibrium_pin,PhysRevLett.110.245702_pin,cammarota2013random_pin,brito2013jamming_pin,gokhale2014growing_pin,nagamanasa2015direct_pin,Szamel}. In the
following, we will show that random pinning allows also 
to improve our knowledge about the density of states in amorphous
solids.

Here, we perform molecular dynamics simulations of a particulate
glass model in three dimensions. After equilibrating the fluid at high temperatures, we compute the density
of the states obtained through the normal modes around the corresponding inherent structures. In computing the inherent structures, we freeze a particle fraction $p$. 
By progressively increasing $p$, we observe that the resulting low-frequency spectrum qualitatively changes. In particular, it develops a non-Debye tail. We find that the low-frequency spectrum is 
populated by soft modes that reach zero as a power law $D(\omega) \sim \omega^\delta$. The exponent $\delta$ departs continuously from
the Debye value $\delta=2$, that is recovered  
at small fraction of pinned particles. For large $p$ values, 
the exponent approaches the limit value $\delta=4$.
Moreover, the non-Goldstone modes become
progressively quasilocalized as the number of frozen particles
increases.


\section*{Results}\label{Results}

\begin{figure}[!t]
\centering
\includegraphics[width=.5\textwidth]{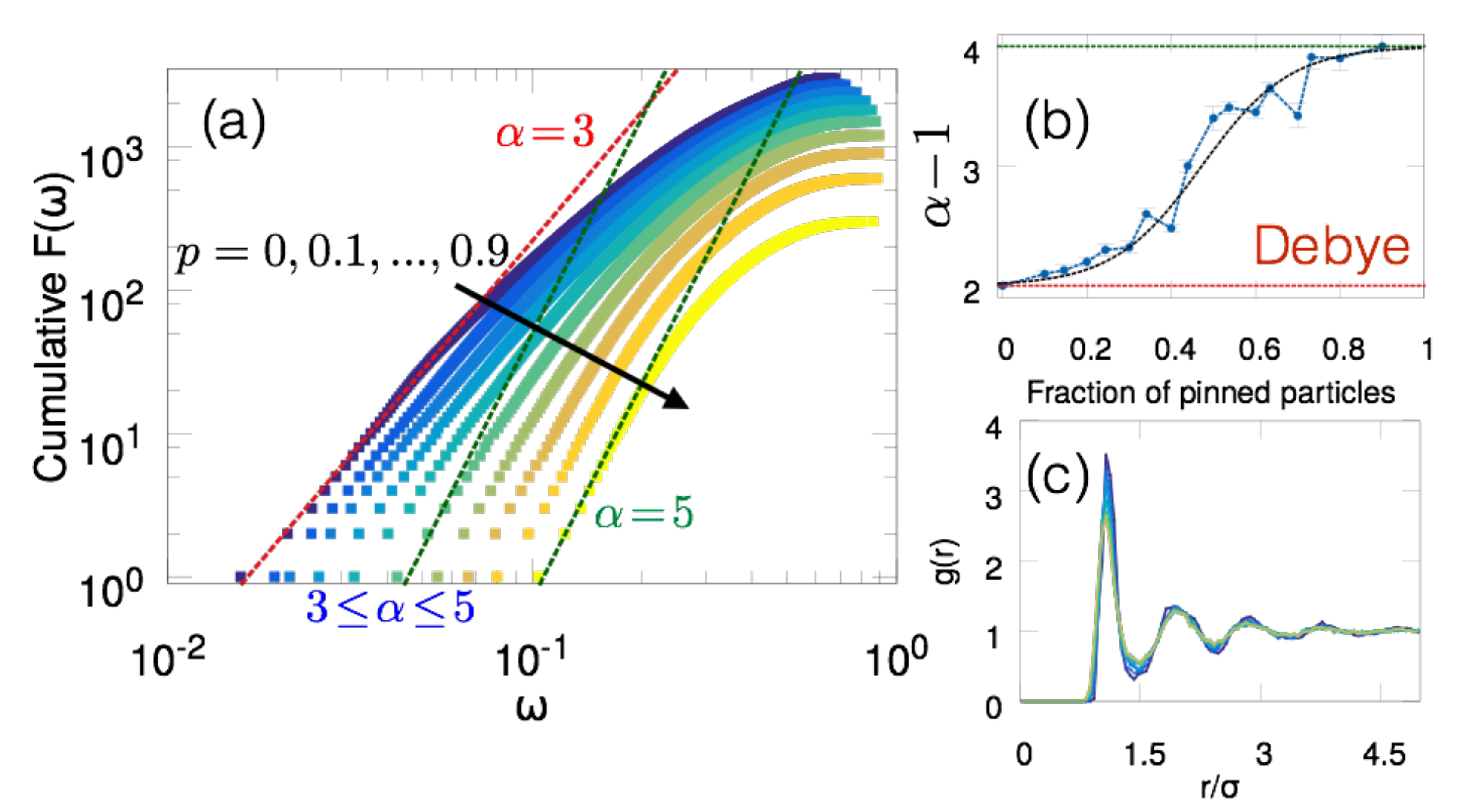}
\caption{ 
(a) The cumulative function {\color{black} of the density of states} 
$F(\omega)$ deviates progressively from Debye by increasing
the fraction of pinned particles $p$, from $0$ (blue) to $0.9$ (yellow). The considered system size is $N=1000$.
(b) From the tail of $F(\omega)$ we fit the exponent $\alpha$ that saturates towards $5$
by increasing $p$. The black dashed curve is the best fit to logistic function. 
(c) Radial distribution function at different pinned particle fraction, from $p=0$ (blue) to $p=0.8$ (yellow).
}\label{fig:fig1}
\end{figure}

\mtt{As a model glass former, we consider a binary mixture $50:50$ composed by
soft-sphere in three dimensions. The particles interact each other through an $r^{-12}$ potential.
The detail about the model and numerical simulations are given in Materials and Methods.
The typical protocol adopted for investigating the vibrational modes are the following.
We start with thermalizing a configuration at a high temperature far above the dynamical
temperature $T_{MC}$ of the model, i. e., the temperature where the system undergoes
a dynamical arrest. The data presented here are obtained working at $T=3\, T_{MC}$.
The inherent structure of the equilibrated configuration is then computed by minimizing
the mechanical energy. In computing the inherent stricture, we randomly choose a finite fraction
$p$ of particles that is maintained frozen during energy minimization. The vibrational modes
are obtained through the diagonalization of the corresponding dynamical matrix.}

We start our discussion by considering a system composed 
by $N=1000$ particles.
Fig. (\ref{fig:fig1})-a shown the cumulative distribution
{\color{black} of the density of states  $F(\omega)$ -- see Materials and Methods.
This quantity shows a power law tail at low frequency, $F(\omega)\sim\omega^{\alpha}$, 
corresponding to a power law tail of the density of states, 
$D(\omega)\sim \omega^\delta=\omega^{\alpha-1}$.}
For $p=0$, the Debye contribution dominates the low-frequency spectrum below the Boson peak. In that region, the cumulative distribution reaches zero as $F(\omega)\sim \omega^{3}$, i. e., $\alpha=d$. 
By increasing the fraction of pinned particles $p$, we observe a progressively disappearing 
of the Goldstone sector in favor of non-Goldstone modes. 
The dashed curves are the fit to the power law $\omega^{\alpha}$, 
we obtain a monotonic growing of $\alpha$ with increasing $p$
that eventually saturates at the value $\alpha = 5$. 
The behavior of $\alpha-1$ as a function of $p$ is shown 
in Fig. (\ref{fig:fig1})-b. In order to give an estimate for the
crossover between Debye and non-Debye regime, we fit the curve $\delta(p)=\alpha(p)-1$
to a generalized logistic curve 
$\delta(p)=\delta_{min} + \frac{\delta_{min} - \delta_{max}}{1+e^{\frac{p-p_{th}}{\sigma}}}$
{\color{black} where $\delta_{min}=2$, $\delta_{max}=4$ and $p_{th}$ and $\sigma$ are the fitting parameters. }
The parameter $p_{th}$ gives an
estimate for the threshold values of $p$ between the two regimes. The logistic curve provides a good
fit to the data with $p_{th}=0.47\pm 0.01$ (black dashed line in Fig. (\ref{fig:fig1})-b).
To be sure that random pinning does not dramatically alterate the structural properties of
the glass, we have kept track of the radial distribution function $g(r)$ of the configuration
that minimizes the energy.  As shown in Fig. (\ref{fig:fig1})-c,
where we plot the evolution of $g(r)$ for different values of $p$, random pinning
does not alterate the structure of the system that remains amorphous even at
large pinned particle fraction.

\begin{figure}[!t]
\centering
\includegraphics[width=.5\textwidth]{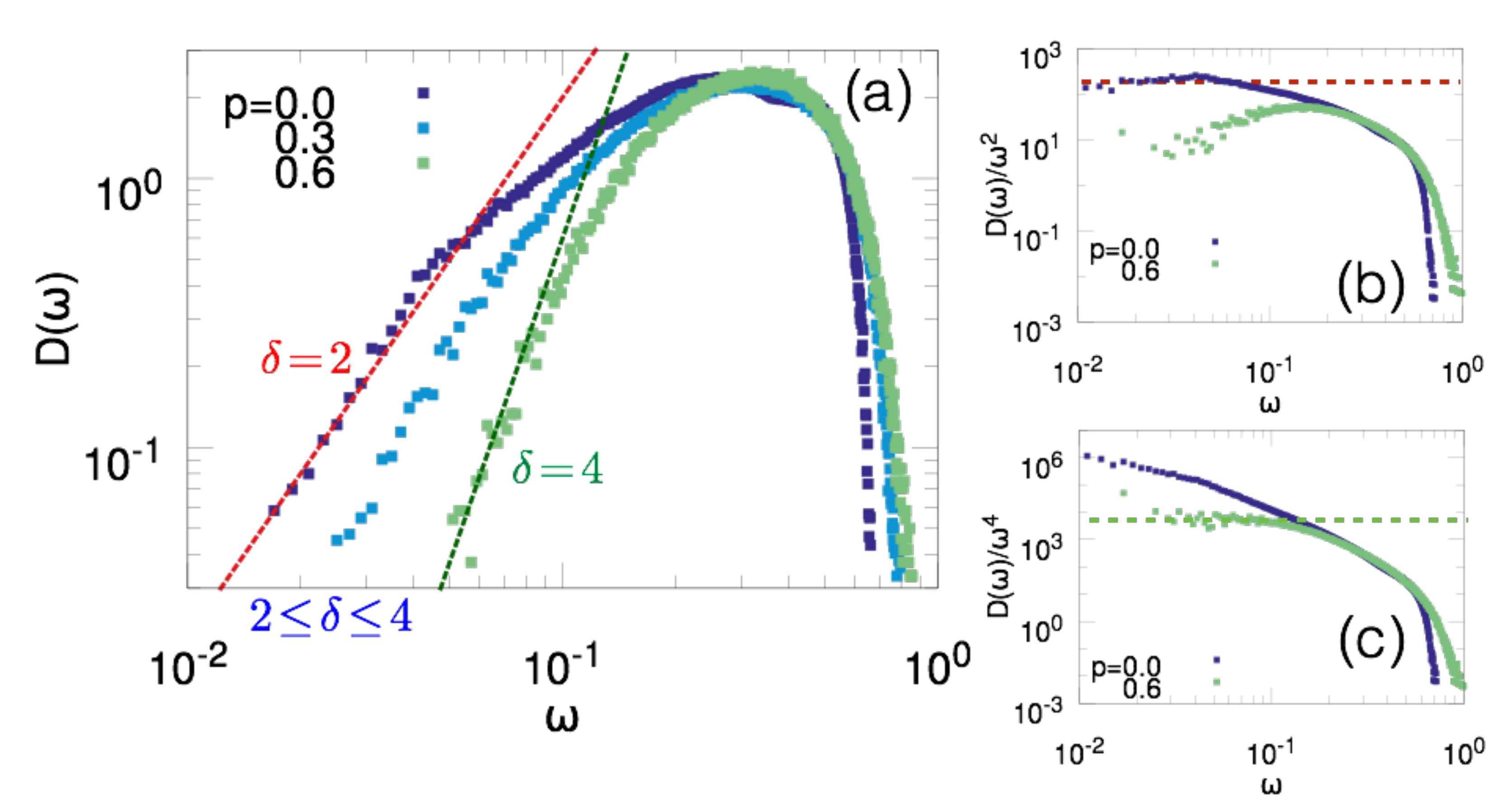}
\caption{ 
(a) The density of states $D(\omega)$
for $N=1000$ particles at $p=0$ (blue), $0.3$ (light-blue), 
and $0.6$ (green). Red dashed curve is Debye's scaling $\sim \omega^2$.
The dashed green curve indicates $D(\omega)\sim \omega^4$.
Debye and non-Debye spectrum are highlighted in
(b) and (c) where we plot $D(\omega)/\omega^2$ and $D(\omega)/\omega^4$,
respectively.
}\label{fig:fig3}
\end{figure}

In Fig. (\ref{fig:fig3}) we show $D(\omega)$ for $N=1000$ particles and $p=0,0.3,0.6$.
Red and green dashed lines in panel (a) are the power laws 
$\omega^2$ and $\omega^4$, respectively.
Of course, at $p=0$, the excess of soft-modes before the low-frequency power
law is the Boson peak. 
The two scaling regimes are made more clear in panels
(b) and (c), where $D(\omega)/\omega^2$ and $D(\omega)/\omega^4$ for
$p=0.0,0.6$ are shown. The dashed lines are a guide to the eye that indicate
$D(\omega)/\omega^\delta=const$.

\begin{figure}[!t]
\centering
\includegraphics[width=.5\textwidth]{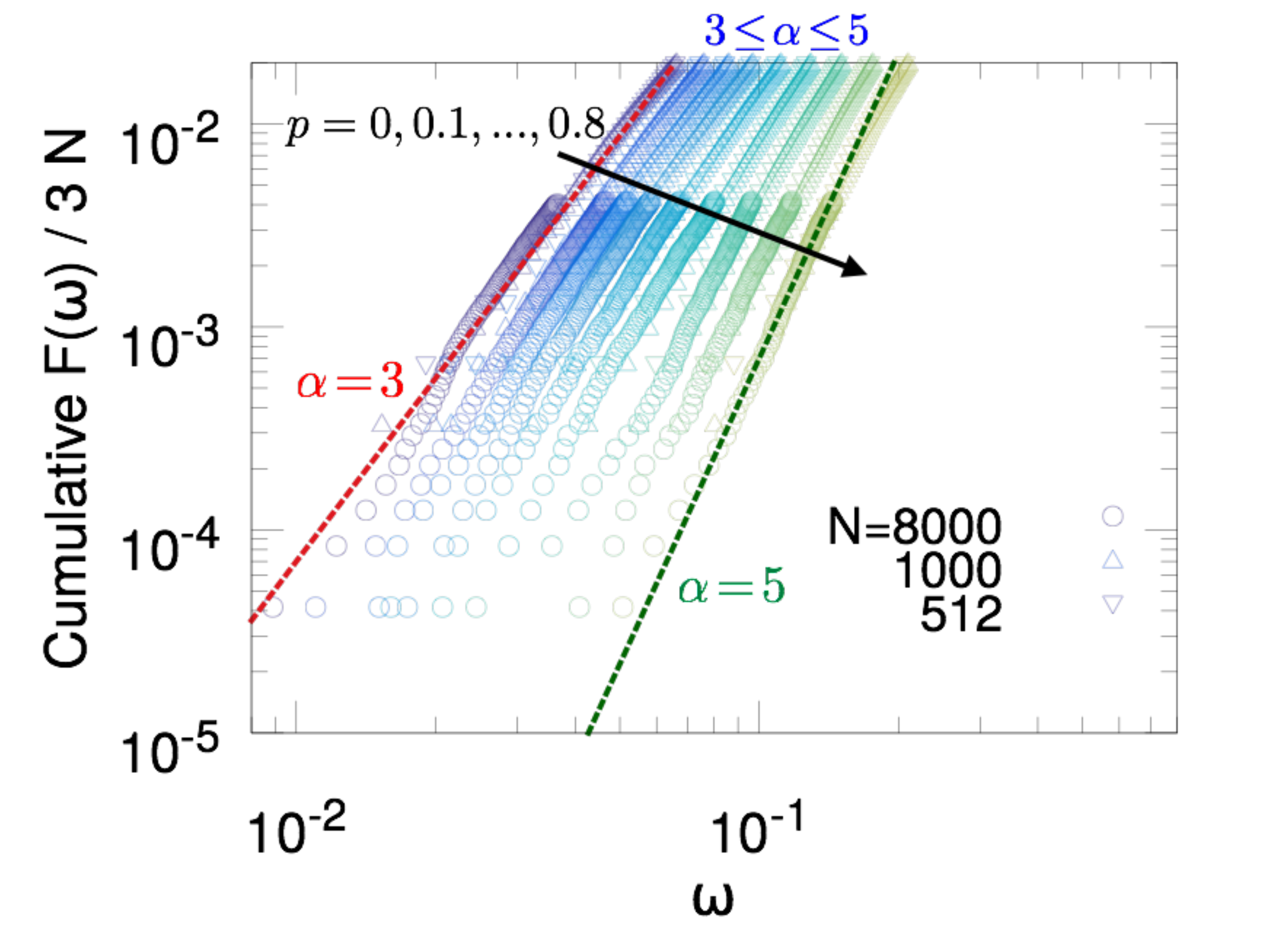}
\caption{ 
Normal modes from different system
sizes $N=512,1000,8000$ smoothly connect with each other. 
Red dashed line and green dashed line are a guide to the eye for
$F(\omega)\sim \omega^3$ and $F(\omega)\sim\omega^5$, respectively.
}\label{fig:fig2}
\end{figure}

In order to take access to lower frequencies, 
we perform simulations up to $N=8000$ particles. The corresponding
$F(\omega)$, opportunely normalized with the total number of modes, is shown in Fig. (\ref{fig:fig2}).
As we can appreciate, 
normal modes from $N=512,1000,8000$ smoothly connect with each other and the corresponding tail
connects continuously with no gap. Again, the exponent of the power law $F(\omega)\sim \omega^{\alpha}$ 
depends on $p$ and interpolates between Debye $ \alpha=3$ and non-Debye $\alpha=5$ as the fraction
of frozen particle increases.
This finding proves that the low-frequency spectrum probed at the smallest size $N=512$
is representative also for larger system sizes.

\begin{figure}[!t]
\centering
\includegraphics[width=.5\textwidth]{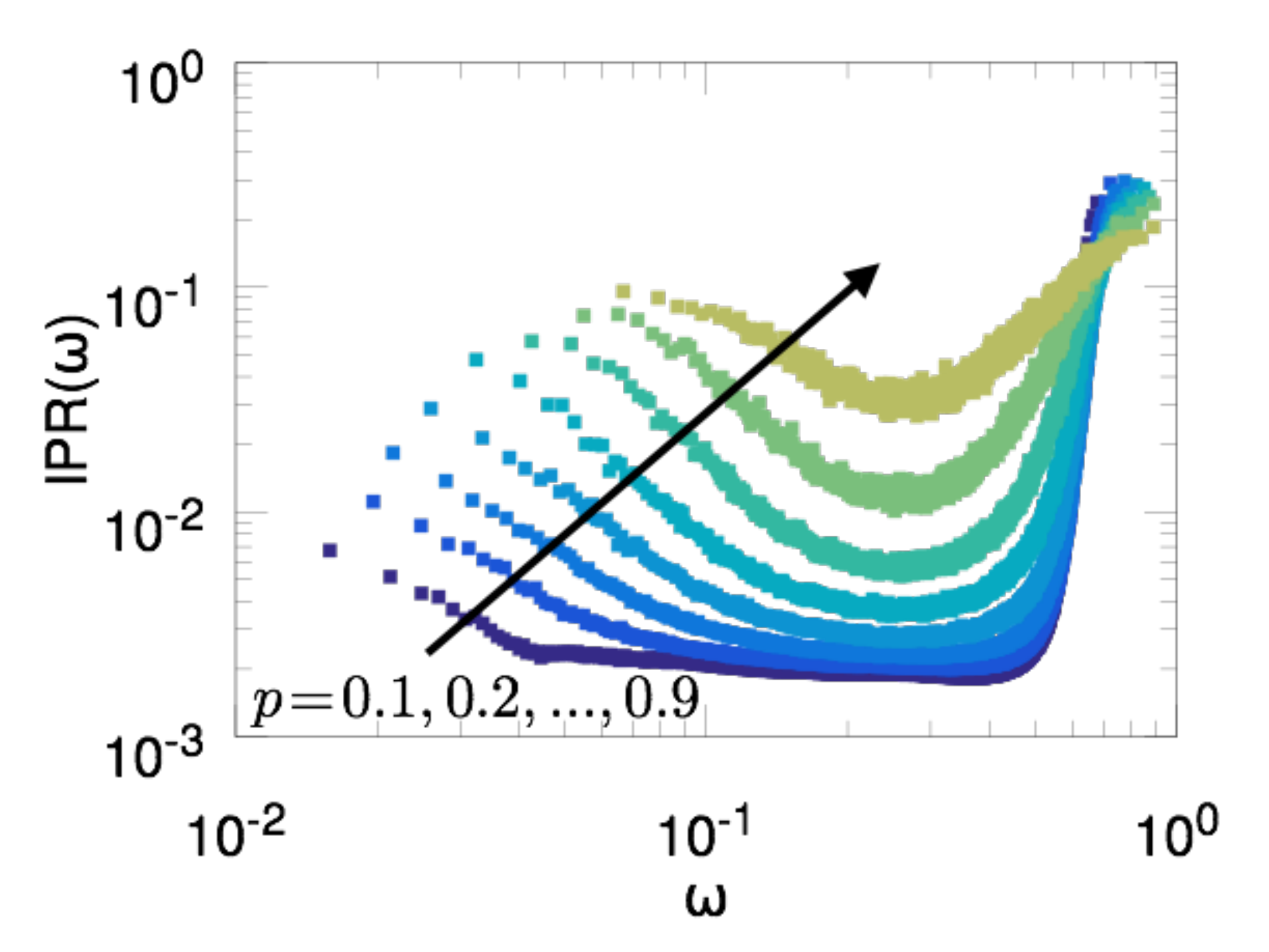}
\caption{ 
Inverse participation ratio 
$IPR$ by increasing the fraction of pinned particles from $p=0$ (purple) to $0.9$ (yellow) 
for a system size $N=1000$.
}\label{fig:fig4}
\end{figure}

Soft modes in glasses are local excitations 
responsible for the plastic flow \cite{degiuli2014forceSoft,Charbonneau_soft,XuVitelliLiuNagelSoft}. Localization can be 
quantified through the inverse participation ratio $IPR(\omega)$.
Delocalized modes cover the entire system and $IPR(\omega)\to 1/N$.
Fully localized modes involve few particles meaning that 
the corresponding eigenvectors count few components. Localization
is then signaled by a scaling $IPR(\omega)\sim 1$. 
The $IPR(\omega)$ computed for $N=1000$ particles is shown in 
Fig. (\ref{fig:fig4}). When $p=0$, phonons dominate the spectrum,
they are extended excitations and, consequently, $IPR\sim 10^{-3}=N^{-1}$.
With increasing the fraction of pinned particles, $IPR(\omega)$ increases too.
Moreover, localization involves particularly the lowest frequency modes. 
This is the signal that modes populating the non-Goldstone sector of $D(\omega)$ 
are progressively quasilocalized in few particle sites.

\section*{Summary and Discussion}
The anomalous thermal and mechanical properties of glasses
are strongly related to their non-Debye excitations
that populate the low-frequency spectrum. For this reason, it is
crucial to develop techniques and protocols that allow to
efficiently probe the sub-dominant glassy modes.

In this paper, we have investigated the low-frequency spectrum 
of a colloidal glass in three dimensions. In order to gain insight into the non-Debye sector of $D(\omega)$, we employ an external field that randomly freezes a particle fraction. 
We showed that the random pinning procedure 
progressively destroys any spatial symmetry presents in
the system removing the phononic contribution to $D(\omega)$.

We have also shown that non-Goldstone modes progressively emerge in the spectrum as the fraction of frozen particles increases. 
Moreover, the power law in the tail of $D(\omega)\sim \omega^{\delta}$ smoothly changes from Debye, i.e., $\delta=2$, to $\delta=4$ at high pinning fraction. The power law 
$ D(\omega)\sim \omega^4$ is fully consistent with the
theoretical prediction for Bosonic non-Goldstone excitations given in Ref. \cite{Gurevich}
that has been also confirmed recently in numerical simulations \cite{Lerner2016,Mizuno14112017,Baity-Jesi15}
{\color{black}  and in the analysis of the sound attenuation in real and simulated glasses \cite{Schirmacher2007} }.
We also found that the low-frequency sector is populated
by soft-modes that become progressively localized 
in few particle sites.

{\color{black} It is worth noting that the scenario here proposed, that is the coexistence at low-frequency of propagating/extended
phononic modes with non-Goldstone/localized states, is consistent with the recently developed "Heterogeneous viscoelasticity"
theory \cite{Schirmacher2015}  where the  role of pinning (or of the "external field") is played by the presence of regions with higher elastic
moduli in the glass.}

As a concluding remark, 
our study suggests an alternative way to probe the properties of
plastic modes in glassy and disordered materials in both, numerical simulations and experiments. 

In computer simulations, 
the protocol considered here has the advantage that works at
any system sizes, at least for the repulsive potentials, i. e., colloidal glasses.
This suggests that, for instance, it is not necessary to tune the system size
below a threshold value as in Ref. \cite{Lerner2016} or
performing large-scale simulations to approach 
the continuum limit \cite{Mizuno14112017}.

Moreover, random pinning fields able to trap 
many colloids through a single laser beam can be
generated in a laboratory by means of speckle patterns \cite{volpe2014speckle} 
{\color{black} or by means of suitable algorithms for the control of spatial light modulator to obtain multi-focus beams
\cite{RDL}}.
In that way, it is possible to isolate the non-Goldstone 
contribution in the low-frequency vibrational modes 
that are accessible in colloidal glasses by means of confocal microscopy \cite{ghosh2010density}.

\matmethods{

\label{Methods}
\subsection*{Model}
As glass-forming model, we consider a $50\!:\!50$ binary mixture of large and small {\color{black} soft spheres} in three dimensions \cite{Grigera01}.
Indicating with $\rr_i$ the position of the particle $i$, with $i=1,...,N$,
two particles $i,j$ interact via a pure repulsive potential $\phi(r_{ij})$, where $r_{ij}\!\equiv \!|\rr_i - \rr_j|$.
The potential reads
\BEQ\label{potential}
\phi(r_{ij})=\left( \frac{\sigma_i + \sigma_j }{r_{ij}}\right)^{12} + \alpha r_{ij} + \beta r_{ij}^2\,,
\EEQ
where $\sigma_i$ takes the value  $\sigma_A$ for the large particles and $\sigma_B$ for the small ones, 
with $\sigma_A/\sigma_B\!=\!1.2$ and $\sigma_A\!+\!\sigma_B\!\equiv\!\sigma\!=\!1$.
We consider $N\!=\!N_A+N_B$ particles {\color{black} ($N_A$=$N_B$)} that are enclosed in a three dimensional box of side $L\!=\!\sigma N^{1/3}$ where periodic 
boundary conditions are employed. The expression for $L$ guaranties $\rho\!=\!N/V\!=\!1$, with $\rho$ the particle density. Moreover, we impose a cutoff to the potential $\phi$ at $r_c\!=\!1.5/(L/2)$ in a way
that $\phi(r)=0$ for $r>r_c$. The coefficient $\alpha$ and $\beta$ are chosen in a way such that $\phi(r)$ has continuous
first and second derivatives at $r=r_c$.

In the following we report all quantities in reduced units.
The MD simulations are performed in $NVT$ ensemble at temperature $T=3\,T_{MC}$, where $T_{MC}$ is
the Mode Coupling temperature of the model, i. e., defined according to $\tau_{\alpha}(T)\sim(T-T_{MC})^{-\gamma}$,
with $\tau_{\alpha}$ the time scale of the $\alpha$ processes. 
We also performed Swap Montecarlo Simulations in $NVT$ ensemble. In that case, the thermalization of the fluid is controlled 
looking at the evolution of the total potential energy $\Phi(\{ \rr_i\})=\sum_{i \leq j} \phi(r_{ij})$ as a function of temperature and comparing it with the Rosenfeld and Tarazona formula $A+B T^{3/5}$ \cite{rosenfeldTarazona1998density}.
The system sizes are $N=512,1000,1024,8000$.

\subsection*{Inherent Structures and Vibrational Mode Analysis}
After thermalisation, we compute the corresponding inherent structures 
that are obtained by minimizing the configurational energy $\Phi(\{ \rr_i\})$.
During the energy minimization, 
we consider a finite particle fraction $pN$, with $p \in[0,1[$,
frozen in its equilibrium configuration. 

In order to compute the density of states, we expand $\Phi$ around the configuration
$\{ \rr_i^0\}$ that minimizes the energy
\BEQ\label{harm_energy}
\Phi(\{ \rr_i\})= \Phi(\{ \rr_i^0\})+\frac{1}{2} \sum_{i, \alpha, j, \beta} \delta r_i^{\alpha} \, \mathcal{H}_{i \alpha,j \beta} \, \delta r_j^\beta + ...
\EEQ  
where $\delta r_i^\alpha \equiv r_i^\alpha - r_i^{\alpha,0}$ {\color{black} and we consider only the non-pinned particles}.
In Eq. (\ref{harm_energy}), we have defined the dynamical matrix
\BEQ\label{hessian}
\mathcal{H}_{i \alpha,j \beta}\equiv \left. \frac{\p^2 \phi(r_{ij})}{\p r_i^\alpha \p r_j^\beta} \right|_{ \{ \rr_i^0\} }\;,
\EEQ
where the latin indices $i,j=1,...,N$ indicate the particles and greek symbols $\alpha,\beta=1,...,d$ the cartesian coordinates.

The energy minimization has been obtained through the Limited-memory Broyden-Fletcher-Goldfarb-Shanno
algorithm \cite{bonnans2006numerical}. For each value of $p$, we 
have collected data for $10^2$ inherent structures obtained
considering different thermalized configuration.
At the end of the minimization, we have checked the structural properties of the corresponding
configuration through the radial distribution function $g(\mathbf{r})=\langle N^{-1} \sum_{i\neq j} \delta(\mathbf{r} - \mathbf{r}_j + \mathbf{r}_i ) \rangle $.

The normal modes are then obtained thorough the diagonalization 
of the dynamical matrix $\mathcal{H}$.
For system sizes up to $1024$, 
we evaluate the entire eigenvalue spectrum through
{\it gsl-GNU libraries}, for larger sizes we compute the lowest $100$ eigenvalues with ARPACK \cite{ARPACK}.
Indicating with $\lambda_\kappa$ the eigenvalues of $\mathcal{H}$, 
the corresponding $3N$ squared normal-mode frequencies are $\omega_\kappa^2=\lambda_\kappa$.
From the spectrum of the eigenvalues $\rho(\lambda)$, using the relation between $\omega_\kappa$ and $\lambda_\kappa$, 
we compute both, the cumulative function $F(\omega)$
and the density of states $D(\omega)$.
The cumulative function $F(\omega)$ reads
\BEQ\label{cumulative}
F(\omega)=\int_0^\omega d\omega^\prime \, D(\omega^\prime) \, .
\EEQ
In the case of a three dimensional elastic solid, the spectrum at low frequencies is populated by phonons, that brings to the power law $F(\omega) \sim \omega^d$ and $D(\omega) \sim \omega^{d-1}$.
If the low-frequency spectrum is populated by non-Goldstone
sof-modes, one expects that $D(\omega)$ reaches zero as
a power law with $D(\omega)\sim \omega^{\delta}$ and,
consequently, $ F(\omega)\sim \omega^{\alpha}$, with $\alpha - 1 = \delta$.

As a measure of the spatial extension of the normal modes,
we compute the inverse participation ratio 
$IPR(\omega)$ defined as \cite{Bell70}
\BEQ\label{IPR}
IPR(\omega) \equiv \frac{\sum_i | \ee_i (\omega )|^4 }{\left( \sum_i | \ee_i (\omega )|^2 \right)^2}
\EEQ
where $\ee_i(\omega)$ is the eigenvector of the mode $\omega$.
For a completely localized mode $\omega$ on a single particle, one has $IPR(\omega)=1$, while a 
mode extended over all the particles corresponds to $IPR(\omega)\sim N^{-1}$.

}

\showmatmethods{} 

\acknow{MP was supported by the Simons Foundation  
Targeted Grant in the Mathematical Modeling of Living Systems Number: 342354 , GP 
and by the Syracuse Soft \& Living Matter Program. This project has received funding from the European Research Council (ERC) under the European Union?s Horizon 2020 research and innovation programme (grant agreement No [694925])}

\showacknow{} 


\bibliography{glassybib}

\begin{thebibliography}{10}

\bibitem{kittel2005introduction}
Kittel C (2005) {\em Introduction to solid state physics}.
\newblock (Wiley).

\bibitem{widmer2008irreversible}
Widmer-Cooper A, Perry H, Harrowell P, Reichman DR (2008) Irreversible
  reorganization in a supercooled liquid originates from localized soft modes.
\newblock {\em Nature Physics} 4(9):711--715.

\bibitem{manning2011vibrational}
Manning ML, Liu AJ (2011) Vibrational modes identify soft spots in a sheared
  disordered packing.
\newblock {\em Physical Review Letters} 107(10):108302.

\bibitem{Hern}
O'Hern CS, Silbert LE, Liu AJ, Nagel SR (2003) Jamming at zero temperature and
  zero applied stress: The epitome of disorder.
\newblock {\em Phys. Rev. E} 68(1):011306.

\bibitem{Wyart12}
Wyart M (2012) Marginal stability constrains force and pair distributions at
  random close packing.
\newblock {\em Phys. Rev. Lett.} 109(12):125502.

\bibitem{Franz15}
Franz S, Parisi G, Urbani P, Zamponi F (2015) Universal spectrum of normal
  modes in low-temperature glasses.
\newblock {\em Proceedings of the National Academy of Sciences}
  112(47):14539--14544.

\bibitem{kurchan2013exact}
Kurchan J, Parisi G, Urbani P, Zamponi F (2013) Exact theory of dense amorphous
  hard spheres in high dimension. ii. the high density regime and the gardner
  transition.
\newblock {\em The Journal of Physical Chemistry B} 117(42):12979--12994.

\bibitem{charbonneau2017glass}
Charbonneau P, Kurchan J, Parisi G, Urbani P, Zamponi F (2017) Glass and
  jamming transitions: From exact results to finite-dimensional descriptions.
\newblock {\em Annual Review of Condensed Matter Physics} 8:265--288.

\bibitem{mezard84}
M\'ezard M, Parisi G, Sourlas N, Toulouse G, Virasoro M (1984) Nature of the
  spin-glass phase.
\newblock {\em Phys. Rev. Lett.} 52(13):1156--1159.

\bibitem{charbonneau2014fractal}
Charbonneau P, Kurchan J, Parisi G, Urbani P, Zamponi F (2014) Fractal free
  energy landscapes in structural glasses.
\newblock {\em arXiv preprint arXiv:1404.6809}.

\bibitem{Gurarie03}
Gurarie V, Chalker JT (2003) Bosonic excitations in random media.
\newblock {\em Phys. Rev. B} 68(13):134207.

\bibitem{Gurevich}
Gurevich VL, Parshin DA, Schober HR (2003) Anharmonicity, vibrational
  instability, and the boson peak in glasses.
\newblock {\em Phys. Rev. B} 67(9):094203.

\bibitem{Schirmacher2007}
Schirmacher W, Ruocco G, Scopigno T (2007) {Acoustic attenuation in glasses and
  its relation with the Boson Peak}.
\newblock {\em Physical Review Letters} 98:025501.

\bibitem{tanaka}
Shintani H, Tanaka H (2008) Universal link between the boson peak and
  transverse phonons in glass.
\newblock {\em Nature materials} 7(11):870.

\bibitem{lerner2016nonlinear}
Lerner E, Gartner L (2016) Nonlinear modes disentangle glassy and goldstone
  modes in structural glasses.
\newblock {\em SciPost Physics} 1(2):016.

\bibitem{Baity-Jesi15}
Baity-Jesi M, Mart\'{\i}n-Mayor V, Parisi G, Perez-Gaviro S (2015) Soft modes,
  localization, and two-level systems in spin glasses.
\newblock {\em Phys. Rev. Lett.} 115(26):267205.

\bibitem{Lerner2016}
Lerner E, D\"uring G, Bouchbinder E (2016) Statistics and properties of
  low-frequency vibrational modes in structural glasses.
\newblock {\em Phys. Rev. Lett.} 117(3):035501.

\bibitem{Lerner_rapid}
Lerner E, Bouchbinder E (2017) Effect of instantaneous and continuous quenches
  on the density of vibrational modes in model glasses.
\newblock {\em Phys. Rev. E} 96(2):020104.

\bibitem{lerner2016statistics}
Lerner E, D{\"u}ring G, Bouchbinder E (2016) Statistics and properties of
  low-frequency vibrational modes in structural glasses.
\newblock {\em Physical review letters} 117(3):035501.

\bibitem{Mizuno14112017}
Mizuno H, Shiba H, Ikeda A (2017) Continuum limit of the vibrational properties
  of amorphous solids.
\newblock {\em Proceedings of the National Academy of Sciences}
  114(46):E9767--E9774.

\bibitem{kob2012non_pin}
Kob W, Rold{\'a}n-Vargas S, Berthier L (2012) Non-monotonic temperature
  evolution of dynamic correlations in glass-forming liquids.
\newblock {\em Nature Physics} 8(2):164.

\bibitem{cammarota2012ideal_pin}
Cammarota C, Biroli G (2012) Ideal glass transitions by random pinning.
\newblock {\em Proceedings of the National Academy of Sciences}
  109(23):8850--8855.

\bibitem{ozawa2015equilibrium_pin}
Ozawa M, Kob W, Ikeda A, Miyazaki K (2015) Equilibrium phase diagram of a
  randomly pinned glass-former.
\newblock {\em Proceedings of the National Academy of Sciences}
  112(22):6914--6919.

\bibitem{PhysRevLett.110.245702_pin}
Kob W, Berthier L (2013) Probing a liquid to glass transition in equilibrium.
\newblock {\em Phys. Rev. Lett.} 110(24):245702.

\bibitem{cammarota2013random_pin}
Cammarota C, Biroli G (2013) Random pinning glass transition: Hallmarks,
  mean-field theory and renormalization group analysis.
\newblock {\em The Journal of chemical physics} 138(12):12A547.

\bibitem{brito2013jamming_pin}
Brito C, Parisi G, Zamponi F (2013) Jamming transition of randomly pinned
  systems.
\newblock {\em Soft Matter} 9(35):8540--8546.

\bibitem{gokhale2014growing_pin}
Gokhale S, Nagamanasa KH, Ganapathy R, Sood A (2014) Growing dynamical
  facilitation on approaching the random pinning colloidal glass transition.
\newblock {\em Nature communications} 5:4685.

\bibitem{nagamanasa2015direct_pin}
Nagamanasa KH, Gokhale S, Sood A, Ganapathy R (2015) Direct measurements of
  growing amorphous order and non-monotonic dynamic correlations in a colloidal
  glass-former.
\newblock {\em Nature Physics} 11(5):403.

\bibitem{Szamel}
Szamel G, Flenner E (2013) Glassy dynamics of partially pinned fluids: An
  alternative mode-coupling approach.
\newblock {\em EPL (Europhysics Letters)} 101(6):66005.

\bibitem{degiuli2014forceSoft}
DeGiuli E, Lerner E, Brito C, Wyart M (2014) Force distribution affects
  vibrational properties in hard-sphere glasses.
\newblock {\em Proceedings of the National Academy of Sciences}
  111(48):17054--17059.

\bibitem{Charbonneau_soft}
Charbonneau P, Corwin EI, Parisi G, Zamponi F (2015) Jamming criticality
  revealed by removing localized buckling excitations.
\newblock {\em Phys. Rev. Lett.} 114(12):125504.

\bibitem{XuVitelliLiuNagelSoft}
Xu N, Vitelli V, Liu AJ, Nagel SR (2010) Anharmonic and quasi-localized
  vibrations in jammed solids—modes for mechanical failure.
\newblock {\em EPL (Europhysics Letters)} 90(5):56001.

\bibitem{Schirmacher2015}
Schirmacher W, Ruocco G, Mazzone V (2015) {Heterogeneous...}
\newblock {\em Physical Review Letters} 115:015901.

\bibitem{volpe2014speckle}
Volpe G, Kurz L, Callegari A, Volpe G, Gigan S (2014) Speckle optical tweezers:
  micromanipulation with random light fields.
\newblock {\em Optics express} 22(15):18159--18167.

\bibitem{RDL}
Di~Leonardo R, Ianni F, Ruocco G (2007) {Computer generation of optimal
  holograms for optical trap arrays}.
\newblock {\em Otpic Express} 15:1913.

\bibitem{ghosh2010density}
Ghosh A, Chikkadi VK, Schall P, Kurchan J, Bonn D (2010) Density of states of
  colloidal glasses.
\newblock {\em Physical review letters} 104(24):248305.

\bibitem{Grigera01}
Grigera TS, Parisi G (2001) Fast monte carlo algorithm for supercooled soft
  spheres.
\newblock {\em Phys. Rev. E} 63(4):045102.

\bibitem{rosenfeldTarazona1998density}
Rosenfeld Y, Tarazona P (1998) Density functional theory and the asymptotic
  high density expansion of the free energy of classical solids and fluids.
\newblock {\em Molecular Physics} 95(2):141--150.

\bibitem{bonnans2006numerical}
Bonnans JF, Gilbert JC, Lemar{\'e}chal C, Sagastiz{\'a}bal CA (2006) {\em
  Numerical optimization: theoretical and practical aspects}.
\newblock (Springer Science \& Business Media).

\bibitem{ARPACK}
Sorensen D, Lehoucq R, Yang C, Maschhoff K (year?) Arnoldi package (arpack).
\newblock {\em www.caam.rice.edu/software/ARPACK/}.

\bibitem{Bell70}
Bell R, Dean P (1970) Atomic vibrations in vitreous silica.
\newblock {\em Discussions of the Faraday society} 50:55--61.

\end{thebibliography}

\end{document}